\documentclass[12pt]{article}
\usepackage{amsmath,amssymb,amsfonts,color,graphicx,cite,color}
\usepackage{latexsym}
\usepackage[normalem]{ulem}
\usepackage{epsfig,psfrag,rotating,soul}
\usepackage{orcidlink}
\usepackage{rotfloat}
\input paperdef

\evensidemargin \oddsidemargin
\allowdisplaybreaks

\begin{document}
\thispagestyle{empty}

\def\thefootnote{\fnsymbol{footnote}}


\vspace{0.5cm}

\begin{center}

\begin{large}
\textbf{Radiative corrections to the $\rm S, T, U$ parameters and their}
\\[2ex]
\textbf{impact on the $W$ boson mass in the 331 model}
\end{large}

\vspace{1cm}

\vspace*{.7cm}

{\sc
M.~Rehman\orcidlink{0000-0002-1069-0637}$^{1}$%
\footnote{email: m.rehman@comsats.edu.pk}%
,~M.A.~Iqbal\orcidlink{0009-0002-3968-801X}$^{1}$%
\footnote{email: mradeeliqbal@yahoo.com}%
,~M.E.~G{\'o}mez\orcidlink{0000-0002-0137-0295}$^{2}$%
\footnote{email: mario.gomez@dfa.uhu.es}%
 and ~O.~Panella\orcidlink{0000-0003-4262-894X}$^{3}$%
\footnote{email: orlando.panella@cern.ch}%

}

\vspace*{.7cm}
{\sl
$^1$Department of Physics, Comsats University Islamabad, 44000
  Islamabad, Pakistan \\[.1em]
$^2$Dpt. de Ciencias Integradas y Centro de Estudios Avanzados en F\'{i}sica Matem\'aticas y Computaci\'on, Campus del Carmen, Universidad de Huelva, Huelva 21071, Spain \\[.1em]
$^3$ INFN, Sezione di Perugia, Via A. Pascoli, I-06123, Perugia, Italy \\[.1em]

}
\end{center}

\vspace*{0.1cm}

\begin{abstract}
\noindent

We investigate radiative corrections to electroweak precision observables, specifically the Peskin--Takeuchi parameters $\mathrm{S}$, $\mathrm{T}$, and $\mathrm{U}$, in the $SU(3)_C \times SU(3)_L \times U(1)_X$ (331) model with $\beta = -\sqrt{3}$. Using the \texttt{SARAH} and \texttt{SPheno} packages, we compute the mass spectrum and low-energy observables. We show that these parameters place strong constraints on the model, requiring most of the scalar masses to lie in the TeV range or below, and imposing indirect bounds on the newly predicted gauge bosons. Furthermore, we demonstrate that the model can accommodate the $W$ boson mass anomaly reported by the CDF collaboration, should future measurements confirm its persistence.

\end{abstract}

\def\thefootnote{\arabic{footnote}}
\setcounter{page}{0}
\setcounter{footnote}{0}

\newpage



\section{Introduction}
\label{sec:intro}

A precise experimental determination of the mass of the $W$ boson can serve as a robust test for the Standard Model (SM). Any observed deviations from the predictions of the SM would provide crucial insights into new physics beyond the SM (BSM). Recently, the Collider Detector at Fermilab (CDF) collaboration announced a new measurement for the mass of the $W$ boson, yielding $M_{W}^{\rm CDF}=80.4335 \pm 0.0094 \gev$~\cite{CDF:2022hxs}. This measurement differs significantly from the SM prediction of $M_{W}^{\rm SM}= 80.353 \pm 0.006 \gev$~\cite{ParticleDataGroup:2024cfk}, with a significance of $7\sigma$. The recent measurements by the CMS \cite{CMS:2024lrd} and ATLAS \cite{ATLAS:2024erm} collaborations are, however, surprisingly close to the SM prediction.  

In addition to the discrepancy in the prediction of the $W$ boson mass reported by the CDF, there are various theoretical and experimental motivations to explore BSM. Numerous models have been proposed to address these concerns. One such model is the 331 extension of the SM, which is based on the concept of higher weak isospin symmetry, denoted by $SU(3)_{L}$, directly extending from $SU(2)_{L}$~\cite{Pisano:1992bxx,Frampton:1992wt,Foot:1992rh,Montero:1992jk,Foot:1994ym}. Notably, this novel gauge framework offers potential solutions to key questions concerning fermion family number~\cite{Pisano:1992bxx}, the quantization of electric charge~\cite{deSousaPires:1998jc, Dong:2005ebq}, discrepancies in the third quark family~\cite{Long:1999ij}, and strong $\cp$ conservation~\cite{Pal:1994ba, Dias:2003zt}. Moreover, it has the potential to address issues such as dark matter~\cite{Mizukoshi:2010ky, Long:2003hht,Neves:2021usp}, flavor physics~\cite{Buras:2014yna, Oliveira:2022vjo, Escalona:2025rxu}, neutrino mass generation~\cite{Tully:2000kk,Doff:2024ovy, Binh:2024lez}, cosmic inflation, and the baryon asymmetry~\cite{Huong:2015dwa,VanDong:2018yae,Dong:2017ayu}.

Various 331 extensions have been proposed in the literature. Among them, the minimal 331 model~\cite{Ferreira:2011hm} and the 331 model with right-hand neutrinos represent two typical examples~\cite{Pisano:1992bxx, Frampton:1992wt, Foot:1992rh}. Additionally, several other intriguing models have emerged, including the economical 331 model\cite{Dong:2006mg}, the reduced 331 model~\cite{Ferreira:2011hm}, the simple 331 model \cite{Dong:2014esa}, the flipped 331 model\cite{Fonseca:2016tbn}, the 331 model with exotic charged leptons~\cite{Pleitez:1992xh}, and the 331 model with neutral (heavy) fermions~\cite{Dong:2010gk}. While certain realizations of the 331 model are known to develop a Landau pole at relatively low energies \cite{Dias:2004dc}, the ~\citeres{Dias:2004wk,Doff:2023bgy} discuss possible ways to avoid this problem.

The particles proposed by the 331 model have not yet been detected in experimental searches. Given the absence of direct experimental evidence, it becomes crucial to scrutinize the indirect effects of these particles on SM observables. For instance, the existence of new particles as predicted by the 331 models may contribute to the Peskin-Takeuchi parameters $\rm S, T, U$~\cite{Peskin:1990zt,Peskin:1991sw,Maksymyk:1993zm}, consequently altering the theoretical prediction of the $W$ boson mass and potentially resolving the $W$ boson mass discrepancy. Furthermore, these effects can be used to indirectly constrain the parameter space of the model, aiding in the evaluation of its feasibility. Some previous work has been done to study the ${\rm S, T, U}$ parameters in the 331 model~\cite{Sasaki:1992np, Frampton:1997in, Long:1999bny, Rodriguez:2022wix}, with the most recent work \cite{VanLoi:2022eir} focusing on the CDF $W$ boson mass discrepancy. However, these previous attempts were mostly generic and did not quantify the impact of the ${\rm S, T, U}$ parameters in terms of model input parameters.

The goal of this research is to scrutinize the indirect effects of particles predicted within the framework of the 331 model on electroweak precision observables (EWPO), particularly focusing on the oblique ${\rm S, T, U}$ parameters. Initially, we will investigate the impact of new particles, specifically the additional Higgs bosons predicted by the 331 model, on the $\rm S, T, U$ parameter. Subsequently, we will analyze the resulting modifications in the $W$ boson mass and identify the appropriate parameter space that can explain the CDF $W$ boson mass discrepancy while adhering to the ${\rm S, T, U}$ constraints.

To achieve this goal, we used the Mathematica package called {\tt SARAH}~\cite{Staub:2009bi,Staub:2010jh,Staub:2012pb,Staub:2013tta,Staub:2015kfa} to create {\tt SPheno}~\cite{Porod:2003um} source code for the 331 model. This allowed us to calculate mass spectra and low-energy observables like the ${\rm S, T, U}$ parameters. We performed the parameter scan using the {\tt SARAH Scan and Plot} (SSP) computer package~\cite{Staub:2011dp}.

The remainder of the paper is structured as follows: In \refse{sec:model_331}, we offer a concise overview of the 331 model, outlining its key components. \refse{sec:CalcSetup} addresses the discrepancy in the $W$ boson mass and a discussion on oblique parameters ${\rm S, T, U}$. We explore how the variation in these parameters influences the shift in the $W$ boson mass. A numerical analysis is presented in \refse{sec:NResults}. Finally, in \refse{sec:conclusions}, we summarize our results and draw relevant conclusions.
\section{Model set-up}
\label{sec:model_331}

The 331 model proposes an extension to the SM with appealing features. It utilizes the gauge group \( SU(3)_C \times SU(3)_L \times U(1)_X \), where \( SU(3)_L \times U(1)_X \) breaks down to \( SU(2)_L \times U(1)_Y \), and subsequently to \( U(1)_Q \), featuring an extended Higgs sector. Anomaly cancellation requirements, along with QCD's asymptotic freedom, dictate the number of generations to match the number of colors, explaining the existence of three generations in the SM. Left-handed fermions transform as triplets under $SU(3)_{L}$, with the number of triplets equaling the number of anti-triplets for an anomaly-free theory.

The 331 model suggests that two quark generations transform as triplets, while one as an anti-triplet, potentially giving an explanation for large top quark mass. Additional heavy quarks are introduced, and in the minimal version, no new leptons are added, with the third component of the lepton anti-triplet chosen as the conjugate of the charged lepton. New heavy neutrinos could also be considered. Moreover, extending $SU(2)_{L}$ to $SU(3)_{L}$ implies the existence of five new gauge bosons, including a neutral boson $Z^{\prime}$. 

The electric charge operator in 331 models is given by:
\[
Q = T_3 + \beta T_8 + X,
\]
where \( T_3 \) and \( T_8 \) are the diagonal generators of \( SU(3)_L \), \( X \) is the \( U(1)_X \) charge, and \( \beta \) is a model-dependent parameter. The choice of \( \beta \) determines the electric charge assignments and the particle content of the model. Common choices include \( \beta =\pm \sqrt{3} \) and  \( \beta = \pm1/\sqrt{3} \). These different $\beta$ values correspond to unique sets of new particles with varying electric charges. Consequently, the particle composition of the 331 model varies according to the specific $\beta$ chosen. In this work, we consider the 331 model with $\beta=-\sqrt{3}$ based on ~\citere{Cao:2015scs}, where newly predicted charged particles are given in \refta{tab:331-Particles} along with their corresponding charges.

\begin{table}[h] \centering
\begin{tabular}
[c]{ccccccccc}\hline\hline
Particles & $D,S$ & $T$ & $E$ & $V$ & $Y$ & $H_{V}$ & $H_{Y}$ & $H_{W}$\\ \hline
$Q $ & 5/4 & -4/3 & -2 & -2 & -1 & -2 & -1 & 1 \\
\hline\hline
\end{tabular}
\caption{New charged particles predicted by 331 model with $\beta=-\sqrt{3}$ along with their respective charges.}%
\label{tab:331-Particles}
\end{table}%
We provide a brief overview of the model in the coming sections.
\subsection{Scalar Sector}

The symmetry breaking pattern of the 331 model is
\[
SU(3)_{L} \times U(1)_{X} \rightarrow SU(2)_{L} \times U(1)_{Y} \rightarrow U(1)_{Q} 
\]
which is accomplished by the introduction of three scalar triplets $\rho $, $\eta $, and $\chi $.

\begin{equation}
\rho =%
\begin{pmatrix}
\rho ^{+} \\ 
\rho ^{0} \\ 
\rho ^{-Q_{V}}%
\end{pmatrix}%
,\text{ }\eta =%
\begin{pmatrix}
\eta ^{0} \\ 
\eta ^{-} \\ 
\eta ^{-Q_{Y}}%
\end{pmatrix}%
,\text{ }\chi =%
\begin{pmatrix}
\chi ^{Q_{_{Y}}} \\ 
\chi ^{Q_{V}} \\ 
\chi ^{0}%
\end{pmatrix}%
\end{equation}
Here $Q_{V}$ and $Q_{Y}$ are the charges on $V$ and $Y$ respectively. The vacuum expectation values (VEVs) of three scalars are chosen as
\begin{equation}
\rho =\frac{1}{\sqrt{2}}%
\begin{pmatrix}
0 \\ 
v_{1} \\ 
0%
\end{pmatrix}%
,\text{ }\eta =\frac{1}{\sqrt{2}}%
\begin{pmatrix}
v_{2} \\ 
0 \\ 
0%
\end{pmatrix}%
,\text{ }\chi =\frac{1}{\sqrt{2}}%
\begin{pmatrix}
0 \\ 
0 \\ 
v_{3}%
\end{pmatrix}%
\label{eta-rho-vev}
\end{equation}

In the first step of symmetry breaking, the field $\chi$ is introduced at a considerably high scale, typically at the $\tev$ scale. This breaking transitions $SU(3)_{L}\times U(1)_{X}$ into $SU(2)_{L}\times U(1)_{Y}$. Subsequently, fields $\rho$ and $\eta$ are employed in the second step of symmetry breaking, which further breaks $SU(2)_{L}\times U(1)_{Y}$ down to $U(1)_{Q}$ at the weak scale, approximately $v _{1}\sim v_{2}\sim M_{W}$. This arrangement ensures $v_{3}$ significantly exceeds $v_{1,2}$.

The scalar potential in the case of scalar triplets mentioned earlier can be expressed as follows:
\begin{eqnarray}
V_{H} &=&\mu _{1}^{2}\left( \rho ^{\dagger }\rho \right) +\mu _{2}^{2}\left(
\eta ^{\dagger }\eta \right) +\mu _{3}^{2}\left( \chi ^{\dagger }\chi
\right) +\lambda _{1}\left( \rho ^{\dagger }\rho \right) ^{2}+\lambda
_{2}\left( \eta ^{\dagger }\eta \right) ^{2}+\lambda _{3}\left( \chi
^{\dagger }\chi \right) ^{2}+\lambda _{12}\left( \rho ^{\dagger }\rho
\right) \left( \eta ^{\dagger }\eta \right)  \nonumber \\
&&+\lambda _{13}\left( \rho ^{\dagger }\rho \right) \left( \chi ^{\dagger
}\chi \right) +\lambda _{23}\left( \eta ^{\dagger }\eta \right) \left( \chi
^{\dagger }\chi \right) +\lambda _{12}\left( \rho ^{\dagger }\eta \right)
\left( \eta ^{\dagger }\rho \right) +\lambda _{13}\left( \rho ^{\dagger
}\chi \right) \left( \chi ^{\dagger }\rho \right)  \nonumber \\
&&+\lambda _{23}\left( \eta ^{\dagger }\chi \right) \left( \chi ^{\dagger
}\eta \right) +\sqrt{2}f\left( \varepsilon _{ijk}\rho ^{i}\eta ^{j}\chi
^{k}+h.c\right) ,
\end{eqnarray}
Here, $f$ has mass dimension, and it is typically assumed to be of the same order as $v_3$ in order to avoid introducing an additional energy scale into the model. The parameters $\lambda_{i}$, $\lambda_{ij}$, and $\lambda^{\prime}_{ij}$ where $i$ and $j$ range from 1 to 3 are dimensionless couplings. 

After the symmetry breaking, we have one $\cp$-odd scalar $A$ and three $\cp$-even scalars denoted by $h$, $H_2$ and $H_3$. For $\cp$-even scalars, the mass matrix is provided as
\begin{equation}
M_{H}^{2}=%
\begin{pmatrix}
2\lambda _{1}v_{1}+\frac{fv_{2}v_{3}}{v_{1}}
& -fv_{3}+v_{1}v_{2}\lambda _{12} & -fv
_{2}+v_{1}v_{3}\lambda _{13} \\ 
-fv_{3}+v_{1}v_{2}\lambda _{12} & 2\lambda
_{2}v_{2}^{2}+\frac{fv_{1}v_{3}}{v_{2}} & 
-fv_{1}+v_{2}v_{3}\lambda _{23} \\ 
-fv_{2}+v_{1}v_{3}\lambda _{13} & -fv
_{1}+v_{2}v_{3}\lambda _{23} & 2\lambda _{3}v
_{3}^{2}+\frac{fv_{1}v_{2}}{v_{3}}%
\end{pmatrix}%
\end{equation}
In the limit $v_{3}\gg v_{1,2}$, the three eigenvalues of the above mass matrix are expressed as
\begin{eqnarray}
M_{h}^{2} &=&\lambda _{1}v_{1}^{2}+\lambda _{2}v_{2}^{2}, \\
M_{H_{2}}^{2} &=&\frac{\left( v_{1}^{2}+v_{2}^{2}\right) }{%
v_{1}v_{2}}fv_{3}, \\
\ M_{H_{3}}^{2} &=&2\lambda _{3}v_{3}^{2}.
\end{eqnarray}

The Higgs boson $h$ is identifiable as the particle detected at the Large Hadron Collider (LHC)\cite{ATLAS:2012yve,CMS:2012qbp} at CERN, with a mass of $125 \, \text{GeV}$. Supposedly, $H_{2}$ and $H_{3}$ are heavier. $h$ and $H_{2}$ emerge from $\rho$ and $\eta$, while $H_{3}$ arises from $\chi$, as a consequence of the two-step symmetry breaking. Additionally, within the Higgs sector of the 331 model, there are three charged scalars denoted as $H_{W}^{\pm}$, $H_{V}^{\pm \pm}$, and $H_{Y}^{\pm}$. The tree-level masses of the Higgs bosons within the 331 model can be summarized as:

\begin{align}
M_{H_{W}}& \approx M_{H_{2}}\approx M_{A}\approx v_{3}\sqrt{\frac{%
k\left( v_{1}^{2}+v_{2}^{2}\right) }{v_{1}v_{2}}} \label{eq:MA-MH2-MHW} \\
M_{H_{3}}& =\sqrt{2\lambda _{3}v_{3}} \label{eq:second} \\
M_{H_{Y}}& =v_{3}\sqrt{k\frac{v_{1}}{v_{2}}+\frac{%
\acute{\lambda}_{23}}{2}} \label{eq:third} \\
M_{H_{V}}& =v_{3}\sqrt{k\frac{v_{2}}{v_{1}}+\frac{%
\acute{\lambda}_{13}}{2}} \label{eq:fourth}
\end{align}
where $k$ is defined by $k \equiv f / v_3$.

\subsection{Gauge Sector}

With the introduction of $SU(3)_L$ symmetry into the gauge framework of the 331 model, four charged gauge bosons ($Y^{\pm}$, $V^{\pm \pm}$) and one neutral gauge boson ($Z^{\prime}$) are required in addition to $W$ and $Z$ bosons. Following the first step of symmetry breaking, these newly introduced gauge bosons obtain their masses, approximately at $v_{3}$. Assuming $v_{3}\gg v_{1,2}$, the mass spectra of the vector bosons can be expressed as

\begin{align}
M_{W}& =\frac{1}{2}g\sqrt{\nu _{1}+\nu _{2}}, \\
M_{Y}& =M_{V}=\frac{1}{2}gv_{3}, \\
M_{Z}& =\frac{1}{\sw}M_{W}, \\
\ \ \ \ \ \ \ M_{Z^{\prime }}& =\frac{2\cw}{\sqrt{3\left[ 1-\left( 1+\beta
^{2}\right) \sw^{2}\right] }}M_{Y},
\end{align}
where $\cw=\rm cos ~\theta _{w}$ and $\sw=\rm sin~\theta _{w}$ and $\theta _{w}$ is the weak mixing angle.

\subsection{Fermion Sector}

Extending the  $SU(2)_L$ gauge group to $SU(3)_L$ leads to left-handed fermions transforming as triplets or anti-triplets under $SU(3)_L$. An anomaly-free theory requires an equal number of triplets and anti-triplets. One approach designates the three lepton generations as anti-triplets, leading to two quark generations as triplets and one as an anti-triplet. Consequently, the first two generations of left-handed quark fields are represented as triplets within the $SU(3)_L$ group, while the third generation is represented as an anti-triplet. This yields the following expressions for the quark fields:

\begin{equation}
q_{_{1L}}=%
\begin{pmatrix}
u \\ 
d \\ 
D%
\end{pmatrix}%
_{L},\ q_{_{2L}}=%
\begin{pmatrix}
c \\ 
s \\ 
S%
\end{pmatrix}%
_{L},\text{ }q_{_{3L}}=%
\begin{pmatrix}
b \\ 
-t \\ 
T%
\end{pmatrix}%
_{L}.
\label{Eq:Quark-Triplets}
\end{equation}
In the \refeq{Eq:Quark-Triplets}, only left-handed quarks are depicted, while their corresponding right-handed counterparts exist as singlets. Furthermore, a different configuration is adopted for the third generation of quarks compared to the SM, driven by the requirement of anomaly cancellation. Additionally, a negative sign is introduced in the $q_{3L}$ triplet to align the $t$ couplings with those of the SM. The lepton fields are represented as

\begin{equation}
l_{_{1L}}=%
\begin{pmatrix}
e \\ 
-\nu _{_{e}} \\ 
E_{_{e}}%
\end{pmatrix}%
_{L},\ l_{_{2L}}=%
\begin{pmatrix}
\mu \\ 
-\nu _{_{\mu }} \\ 
E_{_{\mu }}%
\end{pmatrix}%
_{L},\text{ }l_{_{3L}}=%
\begin{pmatrix}
\tau \\ 
-\nu _{_{\tau }} \\ 
E_{_{\tau }}%
\end{pmatrix}%
_{L}.
\end{equation}

The masses of the new fermions originate from the following Yukawa Lagrangian:
\begin{equation}
    -\mathcal{L}_{\text{Yuk}} = y^J_{ik} \bar{q}_{iL} \chi J_{kR} + y^J_{33} \bar{q}_{3L} \chi^* J_{3R} + y^E_{mn} \bar{\ell}_{mL} \chi^* E_{nR} + \text{h.c.}
\end{equation}
Here, \( i \) (\( k \)) runs from 1 to 2, while \( j \) (\( m, n \)) runs from 1 to 3. The right-handed heavy quarks \( J_{kR} \) correspond to the quarks \( D \) and \( S \), while \( J_{3R} \) refers to \( T \).  
Similarly, \( E_{nR} \) represents the right-handed heavy leptons \( E_e, E_\mu, E_\tau \).  
The scalar field \( \chi \) contributes mass only to the third component of the triplet fermions \( (E_e, E_\mu, E_\tau, D, S, T) \), meaning there is no direct coupling of \( \chi \) to Standard Model (SM) particles.  
Once the first stage of symmetry breaking occurs, the new fermions acquire mass through the vacuum expectation value (VEV) of \( \chi \).  
For simplicity, assuming a diagonal mass matrix for the new fermions, their masses are given by:

\begin{align}
M_{D}& =\frac{y_{11}^{J}}{\sqrt{2}}v_{3},\text{ \ }M_{S}=\frac{y_{22}^{J}}{%
\sqrt{2}}v_{3},\text{ \ }M_{T}=\frac{y_{33}^{J}}{\sqrt{2}}v_{3}, \\
M_{E_{e}}& =\frac{y_{11}^{J}}{\sqrt{2}}v_{3},\text{ \ }M_{E_{\mu }}=\frac{%
y_{22}^{J}}{\sqrt{2}}v_{3},\text{ \ }M_{E_{\tau }}=\frac{y_{33}^{J}}{\sqrt{2}%
}v_{3}
\end{align}

\section{The electroweak precision observables}
\label{sec:CalcSetup}

In the subsequent sections, we offer a concise introduction to the ${\rm S, T, U}$ parameters, followed by an examination of the current experimental status regarding ${\rm S, T, U}$ and the mass of the $W$ boson.

\subsection{The ${\rm S, T, U}$ parameters}

In various BSM models, the dominant BSM effects appear in the corrections to the vacuum polarization of gauge bosons, commonly referred to as oblique corrections\cite{Kennedy:1988rt,Kennedy:1988sn}. If the BSM scale significantly exceeds that of the weak interaction, these corrections can be accurately represented by three distinct parameters: ${\rm S, T, U}$~\cite{Peskin:1990zt,Peskin:1991sw,Maksymyk:1993zm}. These parameters are defined as:

\begin{align}
\hat{\alpha}\left( M _{Z}\right) T& \equiv \frac{\Pi _{WW}^{new}(0)}{M_{W}^{2}}-%
\frac{\Pi _{ZZ}^{new}(0)}{M_{Z}^{2}} \\
\frac{\hat{\alpha}\left( M _{Z}\right) }{4\sw^{2}\cw^{2}}S&
\equiv \frac{\Pi _{ZZ}^{new}(0)}{M_{W}^{2}}-\frac{\Pi _{ZZ}^{new}(0)}{%
M_{Z}^{2}}-\frac{\cw^{2}-\sw^{2}}{\sw \cw}%
\frac{\Pi _{Z\gamma }^{new}(M_{Z}^{2})}{M_{Z}^{2}}-\frac{\Pi _{\gamma \gamma
}^{new}(M_{Z}^{2})}{M_{Z}^{2}} \\
\frac{\hat{\alpha}\left( M _{Z}\right) }{4\sw^{2}}\left( S+U\right) &
\equiv \frac{\Pi _{WW}^{new}(M_{W}^{2})-\Pi _{WW}^{new}(0)}{M_{W}^{2}}-\frac{%
\cw}{\sw}\frac{\Pi _{Z\gamma }^{new}(M_{Z}^{2})}{M_{Z}^{2}}-%
\frac{\Pi _{\gamma \gamma }^{new}(M_{Z}^{2})}{M_{Z}^{2}}
\end{align}
Here, $\hat{\alpha}(M_Z)$ denotes the renormalized electromagnetic fine-structure constant evaluated at the $Z$-boson mass scale, whereas $s_W \equiv \sin\theta_W$ and $c_W \equiv \cos\theta_W$ represent the sine and cosine of the weak mixing angle, respectively. The vacuum polarization functions $\Pi^{\text{new}}_{VV'}(q^2)$ encode the contributions from BSM to the gauge boson two-point functions, with the subscripts $VV'$ referring to specific gauge boson combinations: $\Pi^{\text{new}}_{WW}$ for the $W$ bosons, $\Pi^{\text{new}}_{ZZ}$ for the $Z$ bosons, $\Pi^{\text{new}}_{Z\gamma}$ for the $Z$--photon mixing, and $\Pi^{\text{new}}_{\gamma\gamma}$ for the photon self-energy. The latest values for these parameters\cite{ParticleDataGroup:2024cfk} are given by:

\begin{align}
S& =-0.04\pm 0.10 \\
T& =0.01\pm 0.12 \\
U& =-0.01\pm 0.11
\end{align}%

\begin{figure}[htb!]
\centering
\includegraphics[width=14cm,keepaspectratio]{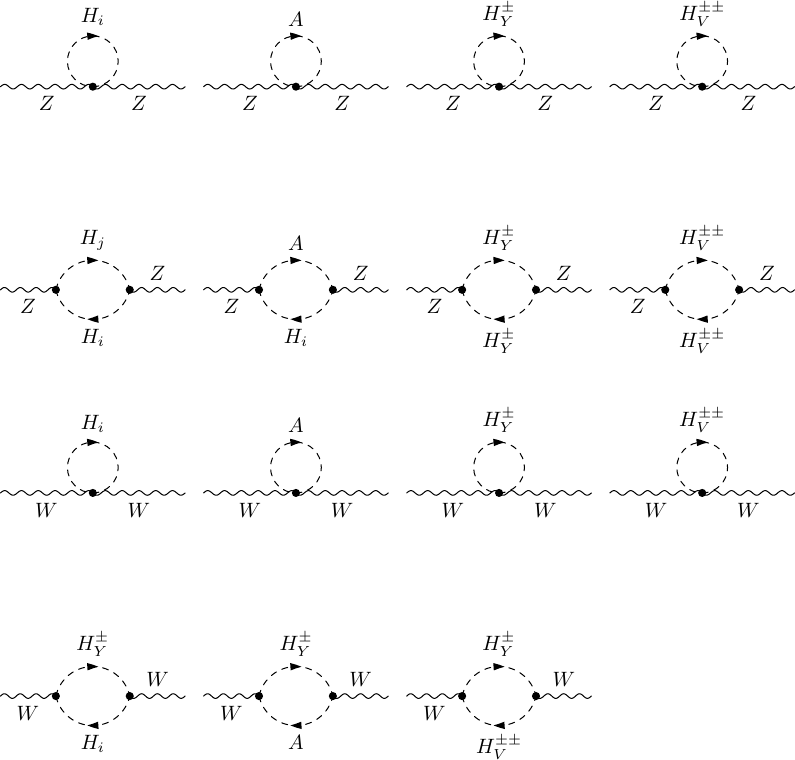}
\caption{Representative Feynman diagrams for the one-loop self-energy of gauge bosons featuring Higgs bosons in the loop, where $H_i$ (with $i=1,2,3$) represent the neutral Higgs bosons.}   
\label{fig:FeynDiag}
\end{figure}

By definition, these values only arise from BSM and align remarkably well with the zero value predicted by the SM. Within the 331 model framework, the introduction of new quarks, which are singlets under $SU(2)_L$, does not influence the oblique corrections to the $\rm S, T, U$ parameters. These corrections are solely sensitive to the breaking of $SU(2)_L$. Similarly, the contribution of the $Z^\prime$ boson is minimal, except through its involvement in $Z-Z^\prime$ mixing. Additionally, the impact of the gauge bosons $Y^{\pm}$ and $V^{\pm \pm}$ on the $\rm S, T, U$ parameters is expected to be negligible\cite{VanLoi:2022eir}. This is attributed to their contribution being contingent upon the mass difference between $Y^{\pm}$ and $V^{\pm \pm}$, which are nearly degenerate. However, dominant contributions are expected from the scalar sector of the 331 model where neutral as well as charged Higgs bosons can play significant role. The representative Feynman diagrams for the scalar bosons are shown in \reffi{fig:FeynDiag} 

\subsection{W boson Mass Discrepancy}

The CDF collaboration has announced a new measurement for the $W$ boson mass, suggesting the potential existence of BSM physics. Their measurement provides a value~\cite{CDF:2022hxs}
\[
M_{W}^{\rm CDF}=80.4335\pm 0.0094~\gev,
\]
showing a notable deviation of around $7\sigma$ from the SM prediction~\cite{ParticleDataGroup:2024cfk}, where
\[
M_{W}^{\rm SM}=80.353\pm 0.006~\gev.
\]
Upon combining previous results from experiments such as ATLAS and LHCb, the world average is determined as~\cite{deBlas:2022hdk}
\begin{equation}
M_{W}^{\rm avg}=80.4133\pm 0.0080~\gev.
\label{CDF-WA}
\end{equation}
However, as the current measurements by the CMS~\cite{CMS:2024lrd} and ATLAS~\cite{ATLAS:2024erm} collaborations are very close to the SM predictions, it has been reported~\cite{LHC-TeVMWWorkingGroup:2023zkn} that a combination of all $W$ boson mass measurements yields a probability of compatibility of only 0.5\%, and is therefore disfavored. A significantly improved compatibility of 91\% is obtained when the CDF-II measurement is excluded from the combination, resulting in the new world average~\cite{ParticleDataGroup:2024cfk}
\[
M_{W}^{\rm avg}=80.3692\pm 0.0133~\gev.
\]

Nevertheless, in this work, we choose to retain the CDF-II result in our analysis. Excluding it would leave little to no room for deviations from the SM, thereby potentially overconstraining BSM scenarios such as the 331 model under consideration. By including the CDF-II result, we explore the potential implications of a higher $W$ boson mass, should this measurement be confirmed, on the parameter space of the model. Therefore, we define $W$ boson mass shift as
\begin{equation}
\Delta M_{W}^{2}= (M_{W}^{\text{CDF}})^2 - (M_{W}^{\text{SM}})^2
= (80.433^2 - 80.357^2)~\gev^2.
\end{equation}

The mass shift of the $W$ boson may be influenced by the $S$, $T$, and $U$ parameters as~\cite{Peskin:1990zt,Peskin:1991sw,Maksymyk:1993zm}
\begin{equation}
\Delta M_{W}^{2} = \frac{\cw^{2}~m_{Z}^{2}}{\cw^{2} - \sw^{2}}%
\left( -\frac{S}{2} + \cw^{2}T - \frac{\cw^{2} - \sw^{2}}{4\sw^{2}}U \right).
\end{equation}

\section{Numerical Results}
\label{sec:NResults}

\subsection{Computational Setup}
\label{sec:comp-setup}

Here's a simplified description of our computational workflow: Using the Mathematica package called {\tt SARAH}~\cite{Staub:2009bi,Staub:2010jh,Staub:2012pb,Staub:2013tta,Staub:2015kfa}, we initially generated the {\tt SPheno}~\cite{Porod:2003um} source code for the 331 model. The source code comprises the analytical expressions essential for computing mass spectra and low-energy observables such as the $\rm S, T, U$ parameters. To conduct parameter scans, we employed an additional Mathematica package called {\tt SARAH Scan and Plot}~\cite{Staub:2011dp}. 

\subsection{Input parameters}
\label{sec:input-para}

The 331 model is characterized by several parameters, including the VEV of the Higgs triplet $\chi$ denoted as $v_3$, a parameter $f$, and dimensionless couplings within the Higgs sector represented by $\lambda_{i}$, $\lambda_{ij}$, and $\lambda^{\prime}_{ij}$ where $i$ and $j$ range from 1 to 3, respectively. Additionally, the model introduces masses for new fermions and defines $\tan\beta$ as the ratio of the vacuum expectation values (VEVs) of the two electroweak-scale scalar triplets, with $\tan\beta \equiv v_2 / v_1$, where $v_2$ and $v_1$ correspond to the VEVs of the neutral components of $\eta$ and $\rho$, respectively, as shown in \refeq{eta-rho-vev}. For the purposes of our numerical analysis, we set the masses of the newly introduced fermions to $800 \gev$, although this is not relevant to our current discussion. As for the remaining parameters, we conducted random scans using the following set:%
\[%
\begin{array}
[c]{ccc}%
 0 \leq& v_{3} & \leq5000 \ \gev\\
 0 \leq& f & \leq 10\ \gev\\
 0 \leq& \lambda_{i}, \lambda_{ij}, \lambda^{\prime}_{ij} & \leq 0.5 \\
0\leq & \tb & \leq 60\ 
\end{array}
\]
We restrict $v_3$ to values up to $5000 \gev$ to avoid the Landau pole\cite{VanLoi:2022eir}. The parameter $f$ is expressed as $f\equiv k v_3$, where typically, the constant $k$ is assumed to be around unity. However, the masses of the Higgs bosons exhibit sensitivity to the magnitude of $f$. This sensitivity leads to large values for the ${\rm S, T, U}$ parameters, consequently imposing constraints on the value of $f$, as will be explained in the following section. Additionally, the mass of the light Higgs boson, $M_h$, is highly sensitive to the input parameters, particularly the couplings $\lambda_1$ and $\lambda_2$. Consequently, we only consider points that result in $M_h$ falling within the range of 123-127 GeV to allow for theoretical uncertainties from unknown higher-order corrections and model-dependent effects.

\subsection{Higgs Boson Masses}
 
\begin{figure}[ht!]
\begin{center}
\psfig{file=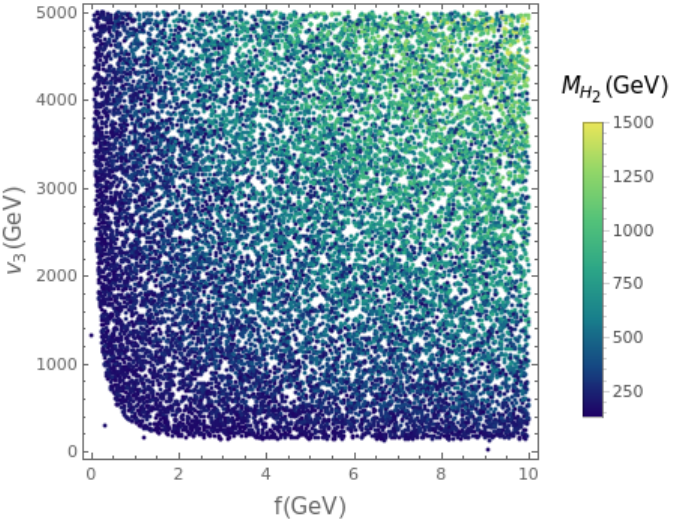  ,scale=0.65,angle=0,clip=}
\hspace{0.5cm}
\psfig{file=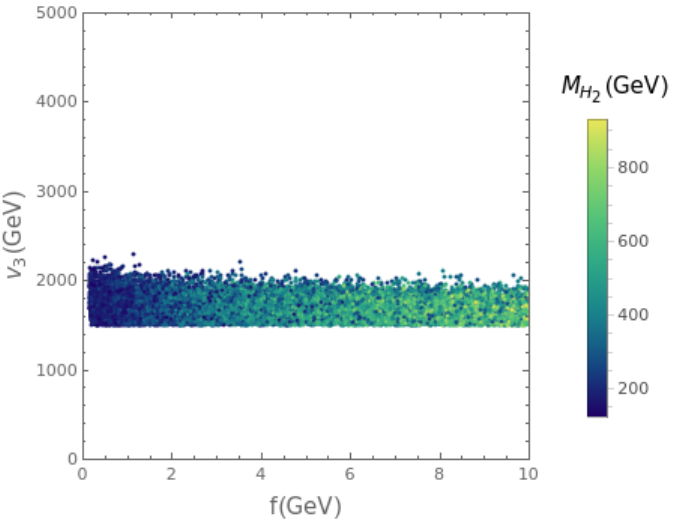  ,scale=0.65,angle=0,clip=}\\
\end{center}
\caption{The  331 predictions for $M_{H_2}$ with (right plot) and without (left plot) ${\rm S, T, U}$ constraints in the $f-v_{3}$ plane. The color bar represents the values of $M_{H_2}$.}
\label{fig:MHH2}
\end{figure}

\begin{figure}[ht!]
\begin{center}
\psfig{file=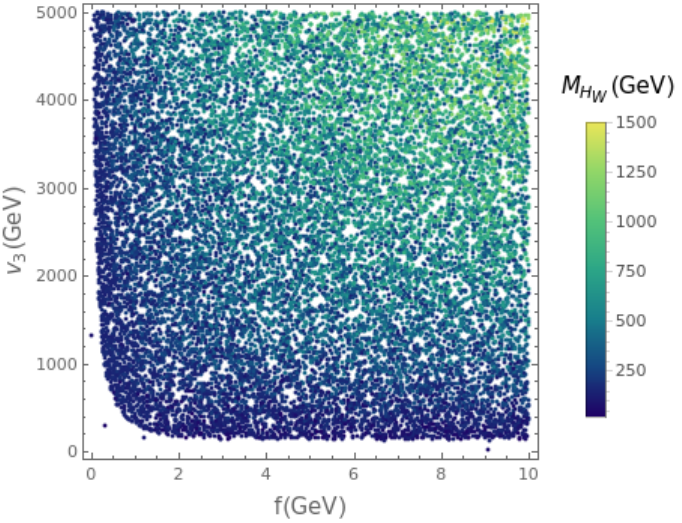  ,scale=0.65,angle=0,clip=}
\hspace{0.5cm}
\psfig{file=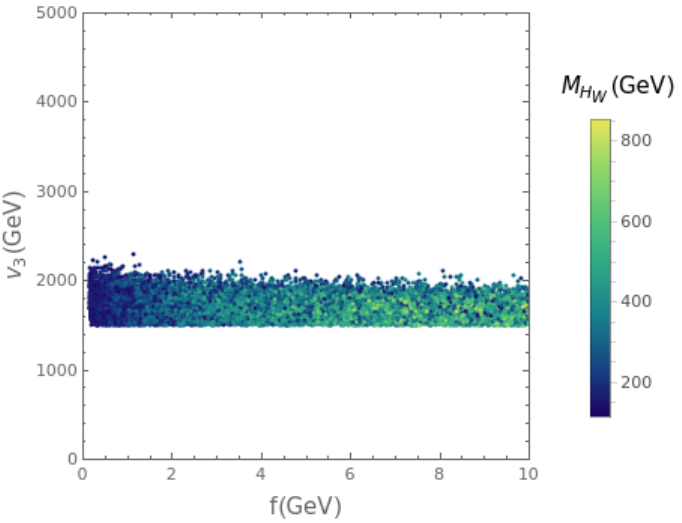  ,scale=0.65,angle=0,clip=}\\
\end{center}
\caption{The  331 predictions for $M_{H_W}$ with (right plot) and without (left plot) ${\rm S, T, U}$ constraints in the $f-v_{3}$ plane. The color bar represents the values of $M_{H_W}$.}
\label{fig:MHW}
\end{figure}

\begin{figure}[ht!]
\begin{center}
\psfig{file=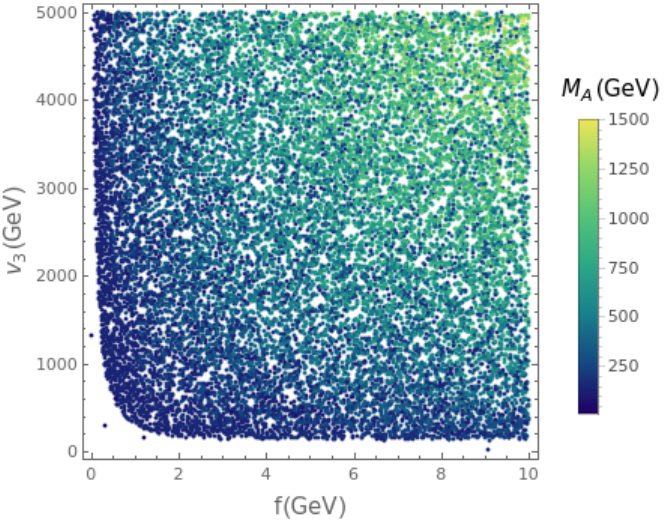  ,scale=0.65,angle=0,clip=}
\hspace{0.5cm}
\psfig{file=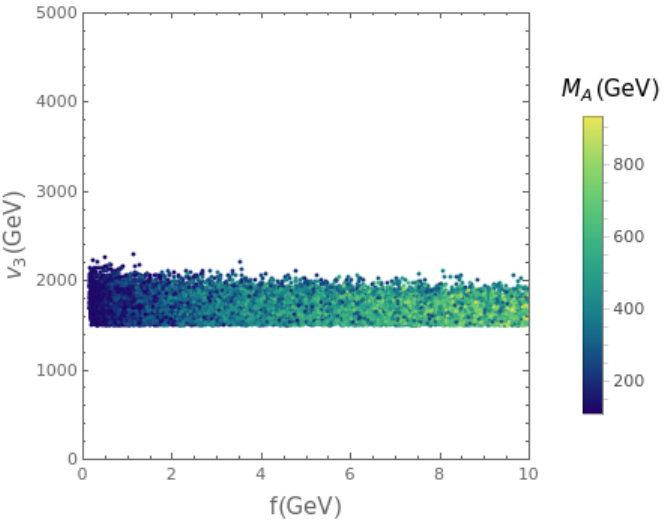  ,scale=0.65,angle=0,clip=}\\
\end{center}
\caption{The  331 predictions for $M_A$ with (right plot) and without (left plot) ${\rm S, T, U}$ constraints in the $f-v_{3}$ plane. The color bar represents the values of $M_{A}$.}
\label{fig:MA}
\end{figure}

\begin{figure}[ht!]
\begin{center}
\psfig{file=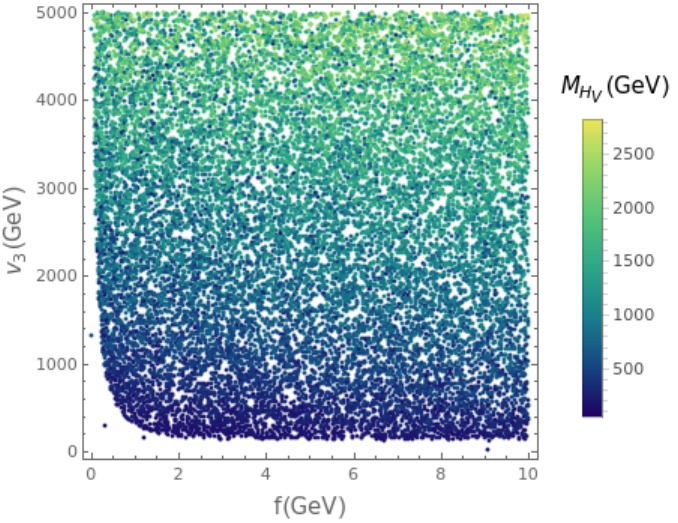  ,scale=0.650,angle=0,clip=}
\hspace{0.5cm}
\psfig{file=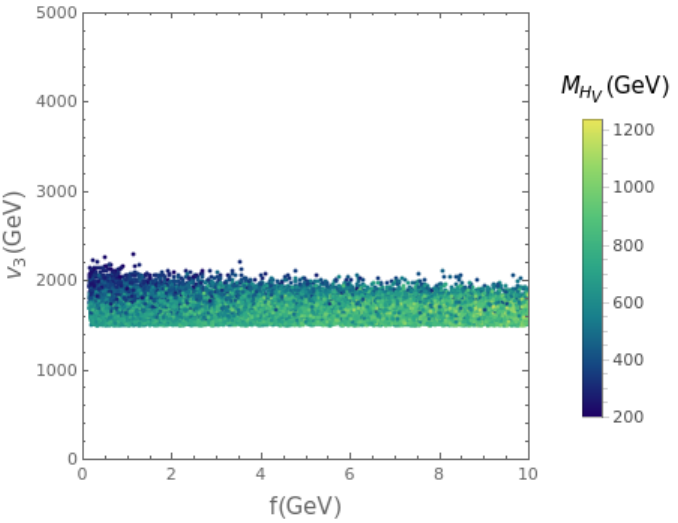  ,scale=0.650,angle=0,clip=}\\
\end{center}
\caption{The  331 predictions for $M_{H_V}$ with (right plot) and without (left plot) ${\rm S, T, U}$ constraints in the $f-v_{3}$ plane. The color bar represents the values of $M_{H_V}$.}
\label{fig:MHV}
\end{figure}

\begin{figure}[ht!]
\begin{center}
\psfig{file=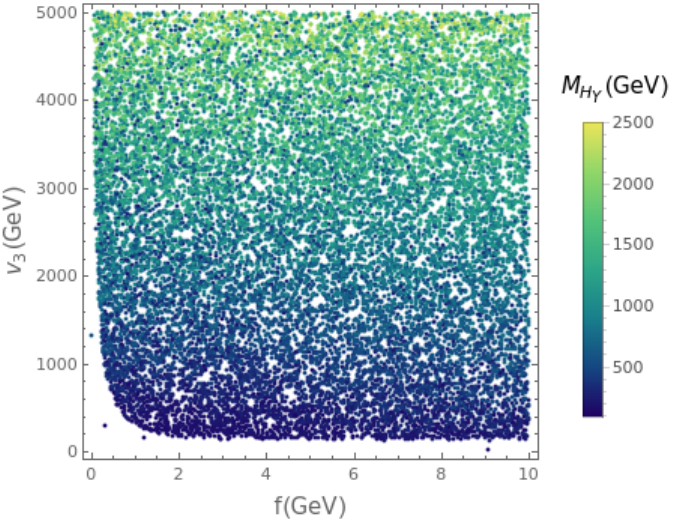  ,scale=0.650,angle=0,clip=}
\hspace{0.5cm}
\psfig{file=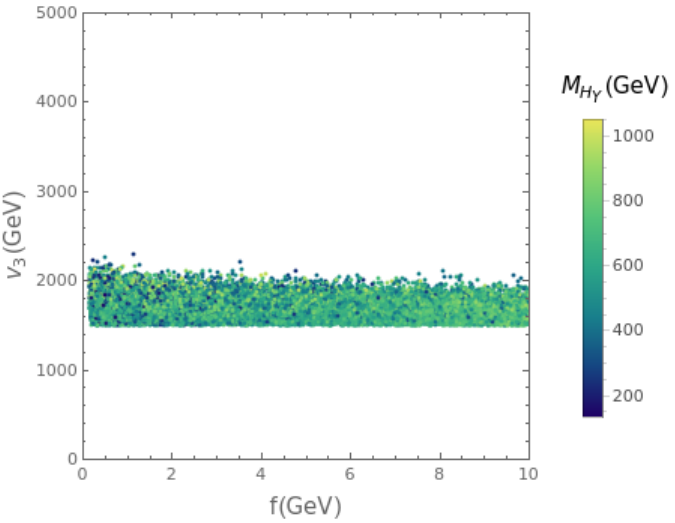  ,scale=0.650,angle=0,clip=}\\
\end{center}
\caption{The  331 predictions for $M_{H_Y}$ with (right plot) and without (left plot) ${\rm S, T, U}$ constraints in the $f-v_{3}$ plane. The color bar represents the values of $M_{H_Y}$.}
\label{fig:MHY}
\end{figure}

\begin{figure}[ht!]
\begin{center}
\psfig{file=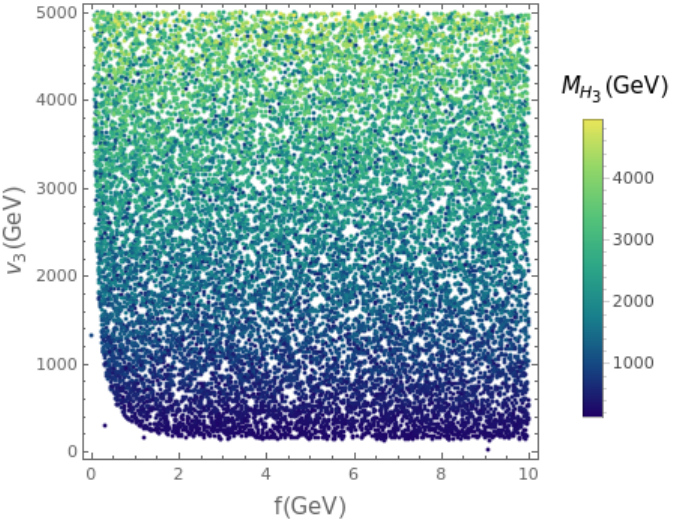  ,scale=0.650,angle=0,clip=}
\hspace{0.5cm}
\psfig{file=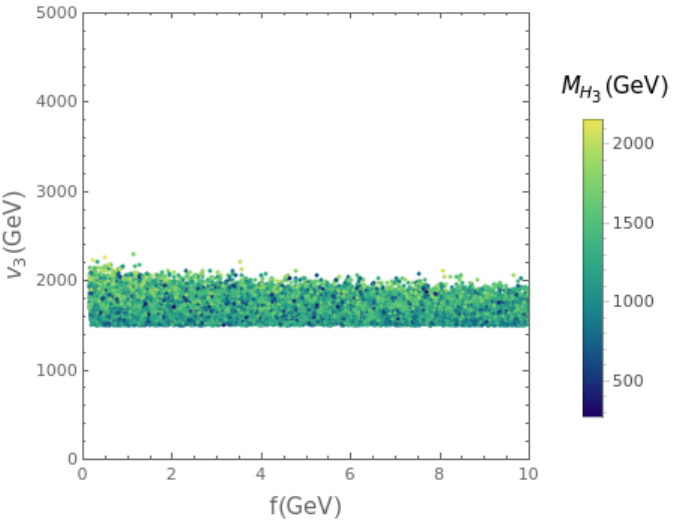  ,scale=0.650,angle=0,clip=}\\
\end{center}
\caption{The  331 predictions for $M_{H_3}$ with (right plot) and without (left plot) ${\rm S, T, U}$ constraints in the $f-v_{3}$ plane. The color bar represents the values of $M_{H_3}$.}
\label{fig:MHH3}
\end{figure}

In this section, we present our results concerning the masses of Higgs bosons plotted on the $f-v_{3}$ plane in \reffis{fig:MHH2}-\ref{fig:MHH3}. Since $M_h$ is independent of $f$ and $v_3$, we do not present the predictions for $M_h$.  \reffi{fig:MHH2} displays the value of $M_{H_{2}}$ in $f-v_{3}$ plane. The left plot showcases our findings without imposing ${\rm S, T, U}$ constraints, while the right plot displays results where only points satisfying ${\rm S, T, U}$ experimental constraints within the $3 \sigma$ range are included. The color bar indicates the magnitude of $M_{H_{2}}$.  As expected, the value of $M_{H_{2}}$ is influenced by both parameters, $f$ and $v_3$. It can be observed that the ${\rm S, T, U}$ constraints preclude values of $v_3$ below $1500 \gev$ and above $2300 \gev$. In the left plot, the dominance of green and yellow points in the region where $v_3 > 2000 \gev$ and $f > 3 \gev$ indicates $M_{H_{2}} > 1000 \gev$. However, the right plot reveals that this region is excluded by the ${\rm S, T, U}$ parameters. As anticipated based on \refeq{eq:MA-MH2-MHW}, $M_{H_{W}}$ and $M_{A}$ exhibit nearly degenerate behavior. Consequently, a comparable trend to that of $M_{H_{2}}$ is observed for $M_{H_{W}}$ and $M_{A}$, as can be seen in \reffi{fig:MHW} and \reffi{fig:MA} respectively.

The correlation of $M_{H_{V}}$ to $f$ and $v_3$ exhibits slight variations compared to $M_{H_{2}}$ and $M_{H_{W}}$. The left plot of \reffi{fig:MHV} illustrates that $M_{H_{V}}$ shows only a mild dependency on $f$, whereas its relationship with $v_3$ is noticeable. Conversely, the right plot in \reffi{fig:MHV} demonstrates that lower values of $M_{H_{V}}$ are favored by the ${\rm S, T, U}$ parameters. \reffi{fig:MHH3} displays the value of $M_{H_{3}}$ in $f-v_{3}$ plane. It indicates that $M_{H_{3}}$ does not exhibit a discernible dependence on $f$, although its reliance on $v_3$ is apparent. Similar conclusions can be drawn for $M_{H_{Y}}$, as evidenced by \reffi{fig:MHY}.


\subsection{Gauge Boson Masses}

\begin{figure}[ht!]
\begin{center}
\psfig{file=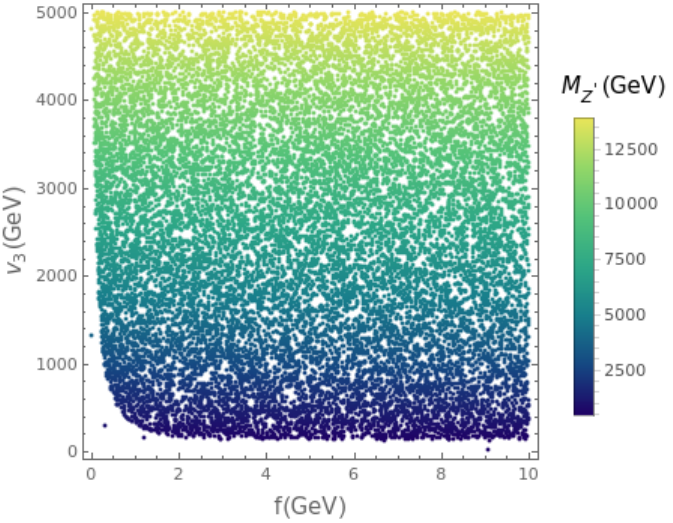  ,scale=0.650,angle=0,clip=}
\hspace{0.5cm}
\psfig{file=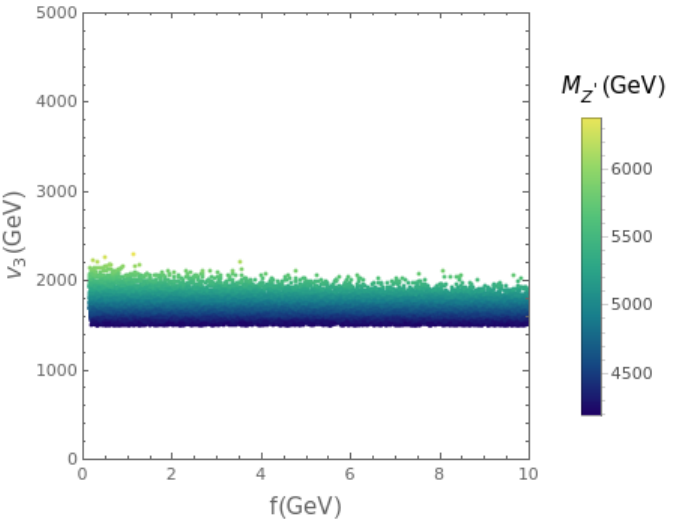  ,scale=0.650,angle=0,clip=}\\
\end{center}
\caption{The 331 predictions for $M_{Z^{\prime}}$ with (right plot) and without (left plot) ${\rm S, T, U}$ constraints in the $f-v_{3}$ plane. The color bar represents the values of $M_{Z^{\prime}}$.}
\label{fig:MZp}
\end{figure}

In this section, we present our results for gauge boson masses, specifically the $M_{Z^{\prime}}$, $M_{Y^{\pm}}$, and $M_{V^{\pm \pm}}$. Although we do not anticipate significant contributions from the $Y^{\pm}$ and $V^{\pm \pm}$ gauge bosons to the ${\rm S, T, U}$ parameters due to their nearly degenerate masses, it is still useful to determine their possible mass ranges considering the influence of the ${\rm S, T, U}$ parameters. In \reffi{fig:MZp}, we present 331 predictions for $M_{Z^{\prime}}$ on the $f-v_{3}$ plane, illustrated in both the right and left plots with and without ${\rm S, T, U}$ constraints, respectively. The color bar indicates the values of $M_{Z^{\prime}}$. It is evident that the ${\rm S, T, U}$ constraints restrict the value of $M_{Z^{\prime}}$ to nearly $7 \tev$, contrasting with its potential range up to $13 \tev$ in their absence.     

The analysis in~\citeres{Coutinho:2013lta, Alves:2022hcp} place lower bounds on \(M_{Z'}\) that can be translated into constraints on the 331 symmetry-breaking scale \(v_3\) via the relation given in Eq.~(15), valid in the decoupling limit \(v_3 \gg v_{1,2}\). The experimental mass limit \(M_{Z'} \gtrsim 4\ \mathrm{TeV}\) quoted there corresponds to the value at which the theoretical prediction for \(\sigma(pp\to Z')\times\mathrm{BR}(Z'\to\ell\ell)\) meets the experimental upper bound. Among other factors, this prediction depends sensitively on the \(Z'\) couplings to quarks and leptons. These couplings control both the production rate, which at LHC energies is dominated by light-quark parton luminosities and scales approximately as \(\sigma(pp\to Z')\propto (g_{Z'}^q)^2\), and the leptonic branching ratio,
\[
\mathrm{BR}(Z'\to\ell\ell)=\frac{\Gamma(Z'\to\ell\ell)}{\Gamma_{\rm tot}},\qquad
\Gamma(Z'\to f\bar f)\propto (g_{Z'}^f)^2 M_{Z'}.
\]
Here, \(g_{Z'}^f\) denotes the effective coupling of the \(Z'\) to a generic fermion \(f\). Any non-uniform change between quark and lepton couplings modifies both the production rate and the leptonic branching fraction, while the opening or closing of additional decay channels alters \(\Gamma_{\rm tot}\) and hence \(\mathrm{BR}(Z'\to\ell\ell)\). Since the fermion charge assignments—and therefore the coupling factors \(g_{Z'}^f\)—differ among 331 variants, in particular between \(\beta=\pm 1/\sqrt{3}\) and \(\beta=-\sqrt{3}\), a faithful translation of the published \(\beta=\pm 1/\sqrt{3}\) limit into our \(\beta=-\sqrt{3}\) setup requires recomputing \(\sigma\times\mathrm{BR}\) with the appropriate charge assignments, while adopting the same acceptance and K-factor assumptions as in the experimental analyses. In practice, the \(\beta= \pm 1/\sqrt{3}\) bound is expected to be somewhat stronger. In the absence of a dedicated recast, and allowing for plausible \(\mathcal{O}(1)\) variations in the couplings and width effects, we conservatively assign a {\bf model-dependent uncertainty} of \(1\!-\!2\ \mathrm{TeV}\), under which the limit remains compatible with our results.

\begin{figure}[ht!]
\begin{center}
\psfig{file=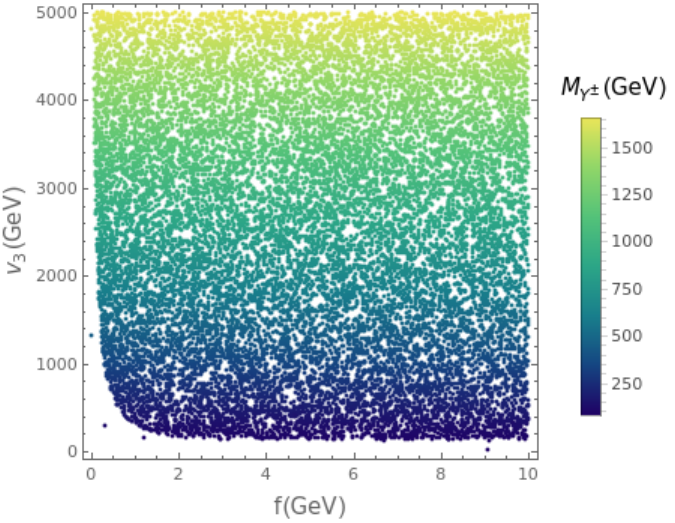  ,scale=0.650,angle=0,clip=}
\hspace{0.5cm}
\psfig{file=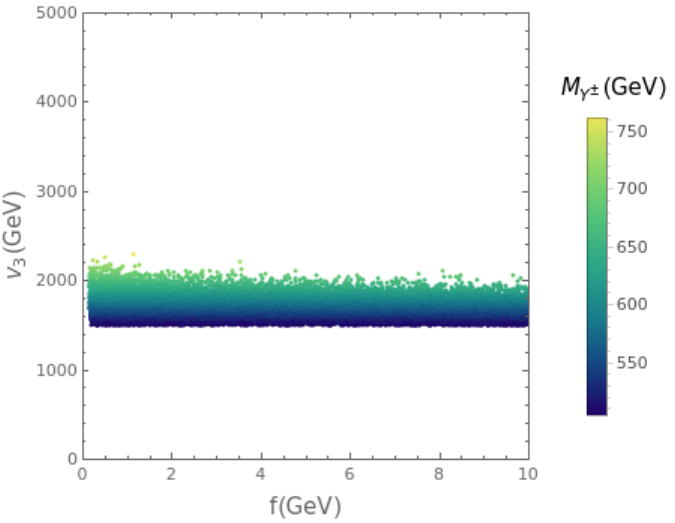  ,scale=0.650,angle=0,clip=}\\
\end{center}
\caption{The  331 predictions for $M_{Y^{\pm}}$ with (right plot) and without (left plot) ${\rm S, T, U}$ constraints in the $f-v_{3}$ plane. The color bar represents the values of $M_{Y^{\pm}}$.}
\label{fig:MYpm}
\end{figure}

\begin{figure}[ht!]
\begin{center}
\psfig{file=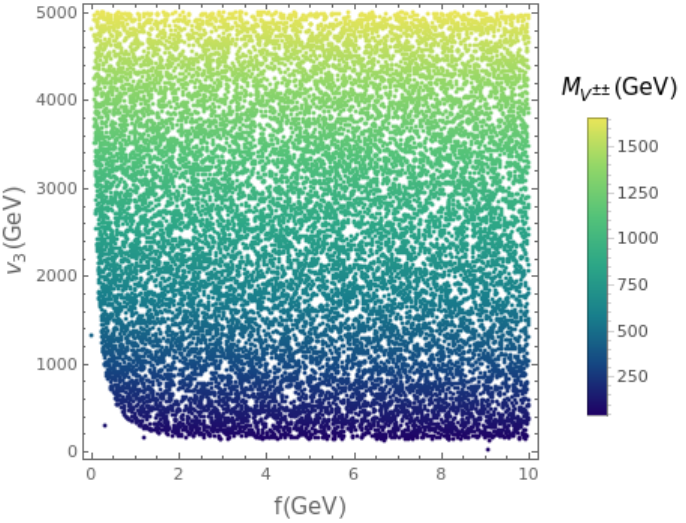  ,scale=0.65,angle=0,clip=}
\hspace{0.5cm}
\psfig{file=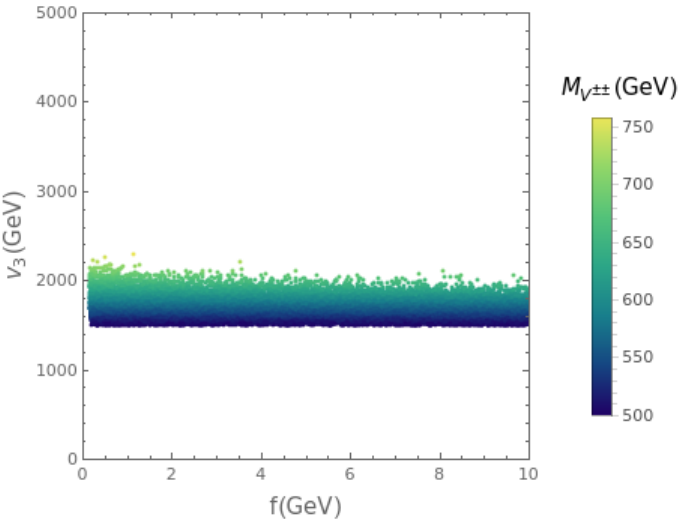  ,scale=0.65,angle=0,clip=}\\
\end{center}
\caption{The 331 predictions for $M_{V^{\pm \pm}}$ with (right plot) and without (left plot) ${\rm S, T, U}$ constraints in the $f-v_{3}$ plane. The color bar represents the values of $M_{V^{\pm \pm}}$.}
\label{fig:MVpmpm}
\end{figure}

Additionally, the values of $M_{Y^{\pm}}$ and $M_{V^{\pm \pm}}$ are similarly bounded to approximately $750 \gev$ by the ${\rm S, T, U}$ constraints, as evidenced by \reffi{fig:MYpm} and \reffi{fig:MVpmpm}. We note that a recent collider recast analysis\cite{Calabrese:2023ryr} of ATLAS multi-lepton data sets a lower limit of about $M_{V^{\pm\pm}} \gtrsim 1.3 \tev$ (90\% C.L.) under the assumption that the doubly charged vector boson decays almost  exclusively into same-sign SM lepton pairs. At face value, this lower bound appears in tension with the upper  bound $M_{V^{\pm\pm}}\lesssim 750 \gev$ inferred from our analysis.  However, several important caveats should be emphasized. First, the collider bound depends sensitively on the  branching ratio into SM leptons: if new decay channels of $V^{\pm\pm}$ into exotic fermions or scalar states  are kinematically open, the effective leptonic branching fraction is reduced and the ATLAS limit correspondingly  weakens. Second, the STU constraints are themselves highly sensitive to the scalar-sector quartic couplings.  While the requirement $M_h \simeq 125$~GeV tightly constrains combinations of $\lambda_1$ and $\lambda_2$, other  quartic couplings (notably $\lambda_3$ and the primed $\lambda'_{i3}$) remain free within our scan and directly control  the heavy-scalar masses and their isospin splittings. Modest variations of these couplings can both reduce the  $T$ parameter by making heavy scalar multiplets more degenerate and open additional cascade decay channels for  $V^{\pm\pm}$. These effects can substantially relax the apparent tension between the STU-derived upper limit  and collider-derived lower bound. A dedicated combined study of oblique parameters, scalar potential parameters, and collider signatures is therefore required before any definitive claims.


\subsection{$\Delta M^{2}_{W}$}

\begin{figure}[ht!]
\begin{center}
\psfig{file=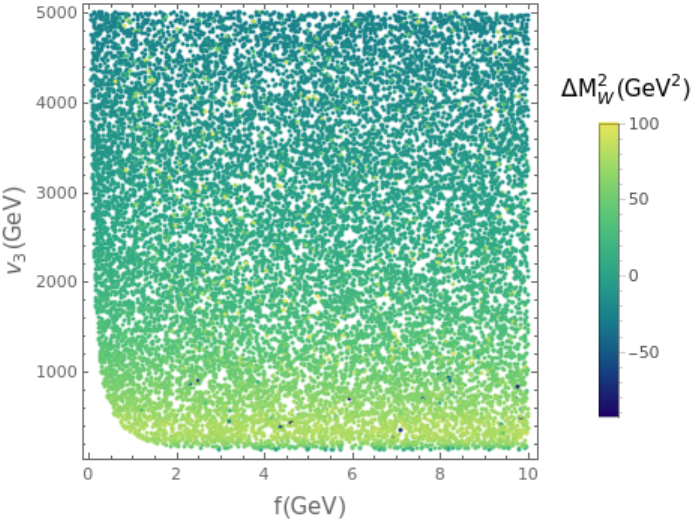  ,scale=0.65,angle=0,clip=}
\hspace{0.5cm}
\psfig{file=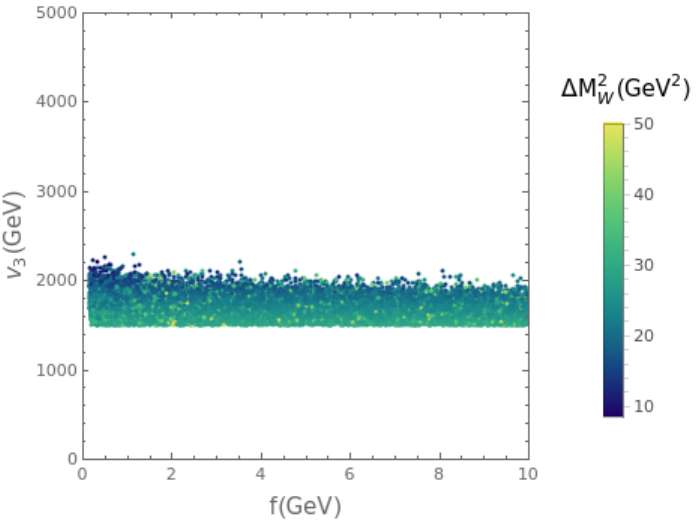  ,scale=0.65,angle=0,clip=}\\
\end{center}
\caption{The  331 predictions for $\Delta M^{2}_{W}$ with (right plot) and without (left plot) ${\rm S, T, U}$ constraints in the $f-v_{3}$ plane. The color bar represents the values of $\Delta M^{2}_{W}$.}
\label{fig:del-MW2}
\end{figure}

\begin{figure}[ht!]
\begin{center}
\psfig{file=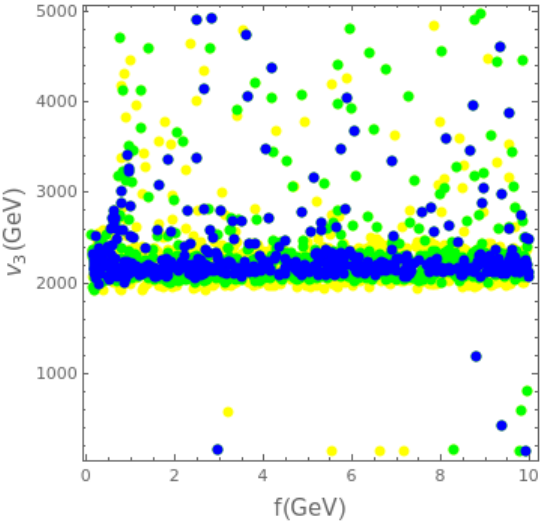  ,scale=0.7,angle=0,clip=}
\hspace{0.5cm}
\psfig{file=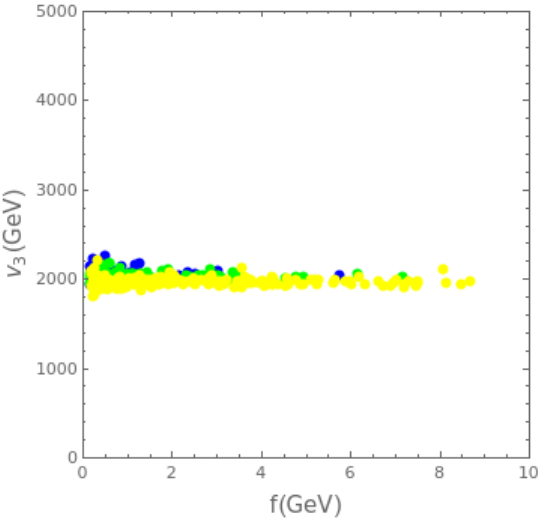  ,scale=0.7,angle=0,clip=}\\
\end{center}
\caption{The  331 predictions for $\Delta M^{2}_{W}$ with (right plot) and without (left plot) ${\rm S, T, U}$ constraints in the $f-v_{3}$ plane. The blue, green, and yellow points denote values of $\Delta M_W^2$ within the $1\sigma$, $2\sigma$, and $3\sigma$ ranges, respectively.}
\label{fig:del-MW2-Points}
\end{figure}

In this section, we present our findings concerning $\Delta M^{2}_{W}$ plotted in the $f-v_{3}$ plane. \reffi{fig:del-MW2} shows the values of $\Delta M^{2}_{W}$ both with (right plot) and without (left plot) adherence to ${\rm S, T, U}$ constraints. Notably, even when considering ${\rm S, T, U}$ constraints, the magnitude of $\Delta M^{2}_{W}$ can be very large, reaching up to $50 \gev^2$. Figure \ref{fig:del-MW2-Points} further explores $\Delta M^{2}_{W}$ in the $f-v_{3}$ plane, focusing specifically on points falling within the $3\sigma$ range of $\Delta M^{2}_{W}$. The left plot disregards ${\rm S, T, U}$ constraints, whereas the right plot exclusively includes points that satisfy ${\rm S, T, U}$ constraints within the $3\sigma$ range. In these plots, blue (green) points denote values within $1\sigma$ ($2\sigma$) of $\Delta M^{2}_{W}$, while yellow points signify values within the $3\sigma$ range.

In the right plot, an interesting trend emerges where the majority of points cluster in the region characterized by small values of $f$. This phenomenon can be attributed to the correlation between the masses of the Higgs bosons, particularly $M_{H_{2}}$, $M_{H_{W}}$, and $M_A$, and the magnitude of $f$. As the magnitude of $f$ increases, the masses of these Higgs bosons also increase, consequently leading to significant contributions to the self-energy diagrams of the gauge bosons involved in the computation of ${\rm S, T, U}$ parameters. Notably, most of the blue points are situated in the region where $f < 3 \gev$. No points satisfying both ${\rm S, T, U}$ constraints and $\Delta M^{2}_{W}$ constraints are observed for $f > 9 \gev$. Furthermore, there are no points outside the range $1800 \gev < v_3 < 2300 \gev$ that meet both the ${\rm S, T, U}$ and $\Delta M^{2}_{W}$ constraints within the $3 \sigma$ range simultaneously.

\section{Conclusions}
\label{sec:conclusions}

The extension of the Standard Model's (SM) gauge structure from $SU(3)_C \times SU(2)_L \times U(1)_Y$ to $SU(3)_C \times SU(3)_L \times U(1)_X$, known as the 331 model, offers a rich phenomenology due to its extended gauge, Higgs, and fermion sectors. Various versions of this model exist which are characterized by a parameter called $\beta$, which determines the types of new particles present. Although no new particles have been detected in experimental searches so far, they could manifest their presence through indirect effects on SM observables, allowing for indirect constraints on the model's parameter space.

Electroweak precision observables (EWPO), particularly the Peskin-Takeuchi parameters ${\rm S, T, U}$ have been measured with high precision and serve as excellent probes for new physics effects. These parameters have the potential to alter the $W$ boson mass predictions, offering a possible resolution to the discrepancy between the SM prediction and the recently reported experimental measurement by the CDF experiment.

In this work, we have investigated the indirect effects of particles predicted by the 331 model, particularly  the additional Higgs bosons, on the Peskin-Takeuchi parameters ${\rm S, T, U}$. Using the {\tt SPheno} setup generated via the Mathematica package {\tt SARAH}, we conducted a parameter scan using \texttt{SARAH Scan and Plot} package as an interface to {\tt SPheno}. Our analysis revealed that the ${\rm S, T, U}$ parameters are highly sensitive to the masses of the Higgs bosons, constraining their values to the $\tev$ range or below. In contrast, the masses of the gauge bosons do not significantly affect the ${\rm S, T, U}$ parameters. However, the ${\rm S, T, U}$ parameters constrain the vacuum expectation value of the third Higgs triplet $v_3$ to the range $1500 \gev < v_3 < 2300 \gev$. This restriction, in turn, imposes upper bounds on the gauge boson masses: $M_{Z^{\prime}} < 7000 \gev$ , $M_{Y^{\pm}} < 700 \gev$, and $M_{V^{\pm \pm}} < 700 \gev$.

We also examined the potential contributions of the ${\rm S, T, U}$ parameters to the $W$ boson mass. We found that the discrepancy in the $W$ boson mass can be successfully explained if the value of $v_3$ lies in the range $1800 \gev < v_3 < 2300 \gev$. The parameter $f$, which appears in the Higgs potential, is typically assumed to be of the same order as $v_3$. However, the ${\rm S, T, U}$ parameters, as well as $\Delta M^{2}_{W}$, favor smaller values for this parameter, restricting its magnitude to $ f < 9 \gev$. While the current study was conducted for $\beta=-\sqrt{3}$, we posit that the results outlined herein are transferable to other 331 versions that share a similar Higgs sector with the one under consideration.  We hope that this work will aid in the experimental searches for 331 particles.

\section*{Declarations}

\noindent\textbf{Funding} \\
M.~E.~G. research is supported by the Spanish MICINN, under grant PID2022-140440NB-C22. 
O.~P. is supported by the INFN grant ENP (Exploring New Physics).

\vspace{1em}
\noindent\textbf{Acknowledgments} \\
M.~R. acknowledges the support and hospitality provided by the University of Huelva and INFN Perugia during the course of this research project.

\vspace{1em}
\noindent\textbf{Conflicts of interest} \\
The authors declare that they have no known competing financial interests or personal relationships that could have appeared to influence the work reported in this paper.

\vspace{1em}
\noindent\textbf{Ethics approval} \\
This article does not contain any studies with human participants or animals performed by any of the authors.

\vspace{1em}
\noindent\textbf{Availability of data and materials} \\
No datasets were generated or analyzed during the current study.

\vspace{1em}
\noindent\textbf{Authors' contributions} \\
All authors contributed equally to the conception, development, and writing of this manuscript. All authors read and approved the final manuscript.


\newpage
\pagebreak

\end{document}